\documentclass[showpacs,twocolumn,prl]{revtex4}
\usepackage{graphicx}
\usepackage{epsfig}
\usepackage{amssymb}

\def\beq{\begin{equation}}
\def\eeq{\end{equation}}
\def\beqa{\begin{eqnarray}}
\def\eeqa{\end{eqnarray}}

\def\e{\epsilon}

\def\half{{\ss 1\over 2}}
\def\third{{\ss 1\over 3}}

\def\D{\Delta}

\def\e{\epsilon}
\def\cH{{\mathcal H}}
\def\cL{{\mathcal L}}
\def\ss{\scriptstyle}

\def\Tr{\mathrm{Tr}}
\def\r{{\bf r}}

\def\d{\mathrm{d}}

\def\ss{\scriptstyle}

\def\etal{{\sl et al.}}

\def\cdag{c^{\dagger}}

\renewcommand{\sec}[1]{\vskip 0.truecm \noindent \emph{#1}. -- }
\newcommand{\bra}[1]{| #1 \rangle}
\newcommand{\ket}[1]{\langle #1 |}

\begin{document}

\title{Theory of non-equilibrium thermoelectric effects in nanoscale junctions}
\author{Y. Dubi and M. Di Ventra}
\affiliation{Department of Physics, University of California San
Diego, La Jolla, California 92093-0319, USA}
\begin{abstract}

%Despite the intrinsic non-equilibrium origin of thermoelectricity in
%nanoscale systems, this phenomenon is usually described within a
%static scattering approach which disregards the dynamical
%interaction with the thermal baths that maintain energy flow. Using
%the theory of open quantum systems we show instead that unexpected
%properties emerge if the non-equilibrium nature of this problem is
%taken into account. We find that the thermopower is a highly
%non-linear function of the temperature gradient, its sign being very
%sensitive to junction geometry. Our approach allows us to define and
%study a local temperature which shows hot spots and oscillations
%along the system according to the coupling of the latter to the
%electrodes. This temperature can be readily accessed experimentally
%and shows that Fourier's law -- a paradigm of statistical mechanics
%-- is violated at the nanoscale.

Despite its intrinsic non-equilibrium origin, thermoelectricity in
nanoscale systems is usually described within a static scattering
approach which disregards the dynamical interaction with the thermal
baths that maintain energy flow. Using the theory of open quantum
systems we show instead that unexpected properties, such as a
resonant structure and large sign sensitivity, emerge if the
non-equilibrium nature of this problem is considered. Our approach
also allows us to define and study a local temperature, which shows
hot spots and oscillations along the system according to the
coupling of the latter to the electrodes. This demonstrates that
Fourier's law -- a paradigm of statistical mechanics -- is generally
violated in nanoscale junctions.

\end{abstract}
\maketitle

Non-equilibrium (NE) processes at the nanoscale are receiving a
great deal of attention due in large part to the advancements in
fabrication and manipulation of these systems.~\cite{Gaspard} An
especially interesting class of NE phenomena pertain to energy
transport and the conversion of thermal to electrical energy. When a
thermal gradient $\Delta T$ is applied to a finite system, electrons
respond by departing from their ground state to partially accumulate
at one end of the system, thus creating a measurable voltage
difference $\Delta V$. The ratio $S=-\frac{\Delta V}{\Delta T}$ is
called thermopower~\cite{Pollock}, and has been measured in a
variety of nano-scale systems such as quantum point contacts
\cite{QPC1}, atomic-size metallic wires \cite{Ruitenbeek}, quantum
dots \cite{QuantumDotsExp}, Si nanowires \cite{Si} and recently in
molecular junctions \cite{Majumdar}. In a bulk material, when $S<0$
the transient current is carried by electrons; when $S>0$ it is
carried by holes.

In nanoscale systems this NE problem has recently received a lot of
attention
~\cite{Datta,QPC1,QPCtheory,QDtheory,MolJuncTHeory,Freericks,Subroto}.
In these theories the single-particle scattering
formalism~\cite{LandauerFormula} is used to relate the thermopower
to single-particle transmission probabilities. This approach,
however, does not take into account the dynamical formation of the
thermopower and neglects the fact that even at steady state, when
the charge current is zero an {\em energy} current is still present,
like, e.g., in insulators \cite{Ashcroft}. Another effect neglected
by such theories, which is now within reach of experimental
verification~\cite{Cahill}, is the formation of local temperature
variations along the structure. In order to study all these effects
one needs to describe a nanoscale system interacting with an
environment that maintains the thermal gradient, namely one needs to
resort to a theory of NE open quantum systems.

In this letter we introduce such a theory, based on a generalization
of quantum master equations, and use it to study the dynamical
formation of thermo-electric effects in nanojunctions. We show that
the thermopower is a highly non-linear function of the thermal
gradient and it is very sensitive to the junction geometry, even in
the simplest case of non-interacting electrons. This precludes an
easy interpretation of its sign in terms of electrons or holes as it
has been argued in some
literature~\cite{Datta,QPC1,QPCtheory,QDtheory,MolJuncTHeory}. In
addition, we calculate the global and local electron distribution
functions, which exhibit NE characteristics.

The theory also allows us to define the local electron temperature
by means of a \emph{temperature floating probe} that is locally
coupled to the system, and whose temperature is adjusted so that the
system dynamics is minimally perturbed. This temperature, which can
be measured experimentally, shows important features such as hot
spots in the cold lead at small coupling between the nanowire and
the bulk electrodes, and temperature oscillations in the wire at
intermediate coupling. These findings show that Fourier's law, which
is considered a paradigm of thermodynamics, is generally violated
for electronic systems at the nanoscale~\cite{Fourier}.

\sec{Method} Since we consider non-interacting electrons coupled to
an environment, we employ a quantum master equation of the Lindblad
type~\cite{lind} which describes the dynamical evolution of the
many-body density matrix (DM) $\rho_M$ of a quantum system in the
presence of a markovian bath, via the introduction of a
super-operator $\cL[\rho_M]$~\cite{vancamp}. The quantum master
equation is then ($\hbar=1$)\beq \dot{\rho}_M = -i[\cH,\rho_M]+\cL [
\rho_M] ~~, \label{lind_eq1} \eeq where $[\cdot,\cdot]$ denotes the
commutator. The super-operator $\cL$ is defined via a set $V_{nn'}$
of operators via
 \beq
 \cL [\rho_M] = \sum_{n,n'} \left( -\half \{ V^\dagger_{nn'} V_{nn'}, \rho_M \}
 +V_{nn'} \rho_M V^\dagger_{nn'} \right) ~~, \label{Lindbladian}
 \eeq
with $\{\cdot,\cdot\}$ being the anti-commutator. The sums over $n$
and $n'$ ($n\neq n'$) are performed over all many-particle levels of
the system, and the $V$-operators are conveniently selected in the
form $V_{n n'}=\sqrt{ \gamma_{nn'}} \bra{\Psi_n}\ket{\Psi_{n'}}$,
describing a transition from the many-body state $\bra{\Psi_{n'}}$
into the state $\bra{\Psi_n}$ with the transition rate
$\gamma_{nn'}$.

This problem scales exponentially with the number of particles, but
recently a mapping of the many-body super-operator to a
single-particle form has been introduced \cite{us}. This results in
a quantum master equation for the {\it single-particle} DM, $
\rho=\sum_{kk'}\rho_{kk'} \bra{k} \ket{k'} \label{singlerho}$ which
provides excellent agreement with the many-body solution. Here,
$\bra{k}$ are the single-particle states, and the matrix elements
are derived from the many-body DM by $ \rho_{kk'}=\Tr \left( \cdag_k
c_{k'} \rho_M\right)$.

\begin{figure}[t]
\vskip 0.5truecm
\includegraphics[width=7truecm]{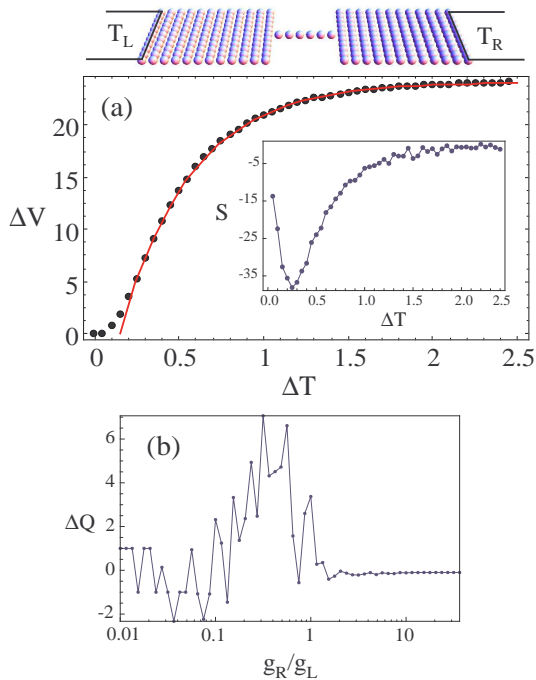}
\caption{\label{dV_dT} (Color online) (a) Upper panel: nanojunction
geometry considered in this calculation (see text for parameters).
Main panel: Voltage drop $\D V$ across the junction as a function of
the temperature difference between the leads. There are three
distinct regimes: the linear regime, a regime of rapid voltage rise
and saturation. Solid line is a fit to an exponential rise. Inset:
generalized thermopower, $S=-\frac{\d (\D V)}{ \d (\D T)}$. (b)
Charge imbalance $\D Q$ as a function of the ratio between the right
and left lead-wire coupling $g_R/g_L$, at a constant $g_L=0.001$ and
at $\D T=1$.}
\end{figure}

To be specific, we consider a finite nano-junction (i.e., with a
fixed number of electrons and ions) which is composed of two
identical quasi-two-dimensional leads connected via a
one-dimensional wire (see upper panel of Fig.~\ref{dV_dT}). The far
edges of the leads are coupled to two different baths kept at
different temperatures. The Hamiltonian of the system is given by $
\cH=\cH_L+\cH_R+\cH_d+\cH_{c} $, where $\cH_{L,R,d}=-t \sum_{\langle
i,j \rangle \in L,R,d} \left(\cdag_i c_j + h.c.\right)$ are the
tight-binding Hamiltonians of the left lead, right lead and wire,
respectively ($t$ is the hopping integral, which serves as the
energy scale hereafter), and $\cH_{c}=
(g_L\cdag_{L}c_{d,0}+g_R\cdag_{R}c_{d,L_d}+h.c. )$ describes the
coupling between the left (right) lead to the wire, with
$\cdag_{L(R)}$ being the creation operator for an electron at the
point of contact between the left (right) lead and the wire, and
$c_{d,0}$ ($c_{d,L_d}$) destroys an electron at the left-most
(right-most) sites of the wire. We consider here spinless electrons.
The master equation now takes the form \beq \dot{\rho} =
-i[\cH,\rho]+\cL_L [\rho]+\cL_R[ \rho]\label{newL}\eeq where
$\cL_{L(R)}$ describes relaxation processes due to the contact
between the left (right) lead with its respective bath at
temperature $T_{L(R)}$. The $V$-operators are generalized to account
for the different baths, and are given by \cite{us,spinchain} \beqa
V^{(L,R)}_{kk'}=\sqrt{\gamma^{(L,R)}_{kk'}f^{(L,R)}_D(\e_k)} \bra{k}
\ket{k'} \label{opeq}~~,\eeqa where $f^{(L,R)}_D(\e_k)=1 /\left(
\exp \left (\frac{\e_k-\mu}{k_BT_{L,R}}\right)+1 \right)$ are the
Fermi distributions of the left and right leads, with $\mu$ the
chemical potential. The coefficients \beq \gamma^{(L,R)}_{kk'}=
\left| \sum_{\r_i\in S_{L,R}} \psi_k(\r_i)\,\gamma_0\,\psi^*
_{k'}(\r_i)\right|\label{gamma} \eeq describe the overlap between
the single-particle states $\bra{k}$ and $\bra{k'}$ over the region
of contact $S_{L(R)}$ between the left (right) baths and the
corresponding junction leads, shown by the solid lines in the upper
panel of Fig.~\ref{dV_dT}. $\gamma_0$ describes the strength of
electron-phonon (bath) interaction. The form~(\ref{gamma}) can be
derived from first principles by tracing out the bath degrees of
freedom, with the latter formed by a dense spectrum of boson
excitations (e.g., phonons), which interact {\em locally} with
electrons at the edges of the system. The operators~(\ref{opeq})
guarantee that the system evolves to a global equilibrium if
$T_L=T_R$, or equilibrate each lead at its own temperature if $g=0$,
i.e., the leads and the wire are completely decoupled (and hence no
voltage drop can form).

We now solve equation~(\ref{newL}) numerically for several
temperature gradients. From the obtained charge density distribution
we derive the electrical potential via the Poisson equation. The
potential is averaged along the transverse direction, and the
voltage drop is calculated from the center of the leads~\cite{Max2}.
The off-diagonal elements of the DM decay fast (on a time-scale
$\sim \gamma^{-1}_0\sim 10$ in our calculations) and hence do not
contribute to the density in the long-time limit. This allows us to
neglect them completely, a fact which significantly simplifies the
calculation~\cite{offdiags}.

\sec{Numerical results} In Fig.~\ref{dV_dT}(a) the voltage drop $\D
V$ across the junction is plotted as a function of the temperature
difference $\D T$ between the contacts. The leads are of dimensions
$12 \times 11$ and the wire is of length $L_d=6$. The lead-wire
coupling is $g_L=g_R=0.001$ and the number of electrons is $n_E=90$,
which corresponds to $\third$ filling. The initial temperatures are
set to $T_L=T_R=0.05$. From Fig.~\ref{dV_dT} one notices three
regimes in the range of $\D T$. At small $\D T$, a linear-response
regime can be identified. This is followed by a regime of rapid rise
in $\D V$, eventually reaching a saturation at large $\D T$, due to
the finite size of the system. The solid line is a fit to an
exponential rise. Although the parameters of the exponential fit
depend on sample parameters, we found that the exponential form is
an excellent fit for all non-interacting junctions. In the inset of
Fig.~\ref{dV_dT}(a) we plot the generalized thermopower,
$S=-\frac{\d (\D V)}{ \d (\D T)}$, which reduces to the regular
thermopower in the linear regime. As seen, $S$ exhibits a resonance
at $\D T\approx 0.25$ (this value is not universal and depends on
junction parameters), which means that at this value the response of
the system to a change in the temperature gradient is maximal; a
fact that can be checked experimentally and may be used in actual
devices.

\sec{Geometrical effects} Due to the local variations of the density
at the junction and hence local variations of kinetic energy, the
thermo-electric response strongly depends on junction geometry, as
it was anticipated experimentally~\cite{Ruitenbeek}. As an example,
we have calculated the charge-imbalance $\D Q$, across a junction
(leads size $6 \times 7$, wire length $L_d=6$, density at $\third$
filling) with an asymmetric coupling between the leads and the wire.
The coupling to the left lead was kept at $g_L=0.001$ and the
coupling to the right lead, $g_R$, was changed. In
Fig.~\ref{dV_dT}(b) we plot the charge imbalance as a function of
$g_R$ at a fixed temperature difference $\D T=1$. Strong and narrow
oscillations can be seen, and for certain values of coupling
asymmetry, $\D Q$ may even change sign. This is consistent with the
experiments in~\cite{Ruitenbeek} and may account for the sign change
of the thermo-power observed in some experiments in molecular
junctions \cite{Ruitenbeek,Majumdar}.

\begin{figure}[t]
\includegraphics[width=9truecm]{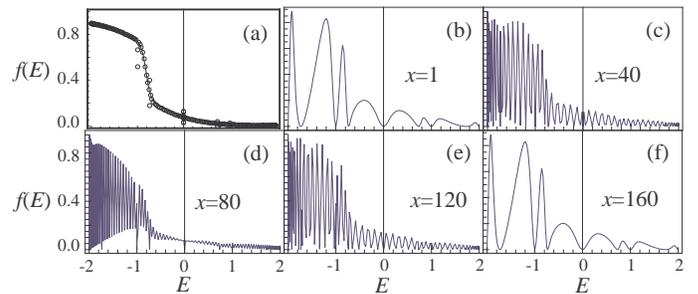}
\caption{ (Color online) (a) Full distribution function of a long
wire, $L_d=160$ (see text for numerical parameters). The solid line
is a mean distribution function $f(E)=\half(f_D(T_L)+f_D(T_R))$.
(b-f) Local distribution function for different positions along the
wire.  } \label{distribution}\end{figure}

\sec{Nonequilibrium distributions} The formalism presented here
allows us to calculate various NE properties. As an example, we
calculate the local and global distribution function (DF) of the
wire (which are accessible experimentally \cite{Pothier}), for a
system with lead size $5 \times 5$, wire length $L_d=160$ (at third
filling), with coupling $g_L=g_R=1$, and temperatures $T_L=0.05$ and
$T_R=0.5$. In Fig.~\ref{distribution}(a) we plot the full
distribution function of the wire (circles), $f(E_k)=\rho_{kk}$,
along with the curve $f(E_k)=\half(f_D(T_L)+f_D(T_R))$ (solid line).
As seen, the fit between the data and the DF is excellent. In
Fig.~\ref{distribution}(b-f) we plot the local DF
$f(E_k,x)=\rho_{kk} |\psi_k(x)|^2$ for different positions along the
wire, $x=1,40,80,120,160$ (respectively). Note the fast oscillations
upon approaching the center of the wire, and the symmetry with
respect to the wire center. The origin of the oscillations lies both
in the NE nature of the system and in the geometry (i.e. the fact
that the wire is ballistic and that the wire-lead coupling is
large). In a system with small wire-lead coupling we found (not
shown) that the oscillations persist at the center of the wire, but
are smoothed at the wire edges. Fig.~\ref{distribution}(b-f) shows
that although the full DF is a simple average of the left and right
lead distributions, one cannot assign a simple position dependence,
as in the case of mesoscopic wires \cite{Pothier}.

\begin{figure}[t]
\vskip 0.5truecm
\includegraphics[width=9truecm]{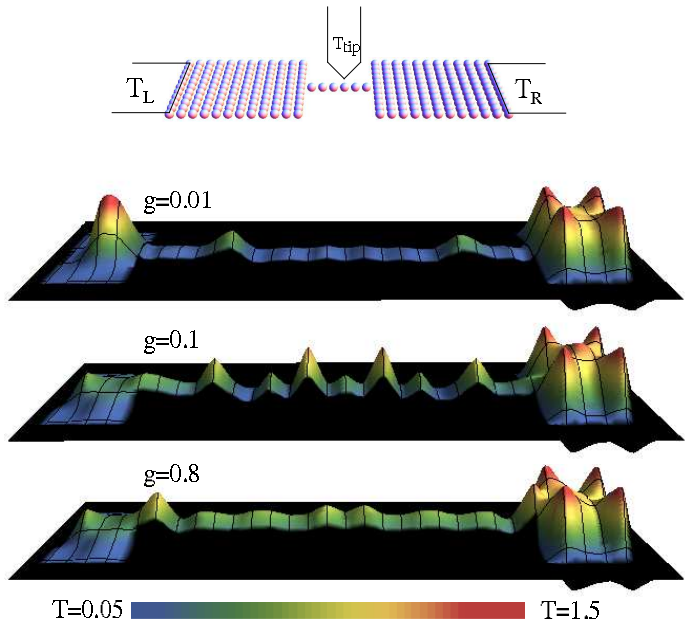}
\caption{ (Color online) Upper panel: schematic representation of
the calculation to determine the local temperature via the addition
of a local temperature floating probe. Lower panels: the local
temperature along the structure for three different values of the
lead-wire coupling, $ g=0.001,0.1,0.8$. Three effects are observed:
a hot-spot in the cold lead at small coupling, temperature
oscillations in the wire at intermediate coupling, and uniform
temperature along the wire at large coupling. All three constitute
violations of Fourier's law. } \label{T_local}\end{figure}

\sec{Local temperature} The concept of a local temperature is
generally not unique out of equilibrium~\cite{localT}. Here we
provide an operational definition, which is both physically
transparent and can (in principle) be directly probed
experimentally~\cite{Cahill}. In order to do so, we add an
additional relaxation operator $\cL_{{\rm tip}} [ \rho_M]$ to the
master equation~(\ref{lind_eq1}). This corresponds to applying a
local bath at a temperature $T_{{\rm tip}}$ in contact with a single
site of the system (see upper panel of Fig.~\ref{T_local}). Due to
energy flow between the probe and the coupled site, the system
dynamics is generally modified. However, one can ``float'' the
temperature $T_{{\rm tip}}$  such that the change in local (and
hence global, e.g., thermopower) properties of the system is
minimal. We define this temperature as the {\em local temperature}
of our system at the probe position. We choose to monitor the change
in the local density from its value in the absence of the probe, but
any other quantity would be equally valid and lead to the same local
temperature.

Knowledge of the local temperature allows us to address the problem
of validity of Fourier's law at the nano-scale. It has been
conjectured \cite{Fourier} and demonstrated for systems such as
spin-chains \cite{spinchain,Michel} and coupled harmonic oscillators
\cite{harmonicchain} that in ballistic quantum systems Fourier's law
is invalid. Stimulated by this open problem, we plot in
Fig.~\ref{T_local} the local temperature at steady state for three
different values of the lead-wire coupling, $g=0.001,0.1,0.8$ and a
junction with lead dimensions $4\times 3$, wire length $L_d=21$,
temperatures $T_L=0.05$ and $T_R=1.5$, and electron density at third
filling. For weak coupling ($g=0.001$), we find that the temperature
inside the wire is very low, but a ``hot spot'' develops in the cold
lead. As the coupling increases the hot spot vanishes, and
temperature oscillations develop in the wire. As expected, at high
coupling, the wire equilibrates at a temperature that is roughly the
average between $T_L$ and $T_R$. For large lead-wire coupling the
temperature in the wire is uniform, and most of the temperature drop
occurs at the contacts, similar to what has been argued for the
phenomenon of local ionic heating~\cite{Chen}. We note that when the
coupling $g \sim 1$, so that the wave-functions are completely
delocalized along the nano-junction, the thermopower is small. This
is consistent with the experimental results of~\cite{Ruitenbeek},
where junctions of large conductance show small thermopower values.
Our results verify that Fourier's law is generally violated in
nanoscale junctions.

\sec{Summary} We have shown, using an open quantum system approach,
that the proper description of energy flow in nanoscale systems
leads to unexpected features in the dynamical formation of
thermopower. Several predictions have been made which can be tested
experimentally, namely (i) the non-linear dependence (with resonant
structures) of the thermopower as a function of the temperature
gradient, (ii) its strong sensitivity (including sign) to junction
geometry, (iii) the shape of the NE electron distribution at steady
state, and (iv) we have provided an operational definition of local
temperature as that measured by a temperature floating probe, i.e.,
one that is locally coupled to the system, and whose temperature is
adjusted so that the system dynamics is minimally perturbed. This
temperature shows noteworthy features according to the strength of
the coupling between the nanoscale system and the electrodes, in
violation of Fourier's law.

We conclude by pointing that the results presented here may be
relevant to other systems of present interest (e.g. graphene
nano-ribbons, nanotubes etc.). Studies of these effects that include
electron interactions represent another important research
direction, and are underway using stochastic time-dependent current
density-functional theory \cite{STDCDTF}.

We thank R. D'Agosta for useful discussions and DOE for support
under grant DE-FG02-05ER46204.


\begin{thebibliography}{}

\bibitem{Gaspard}
For a review see, e.g., P. Gaspard, Prog. Theor. Phys. Suppl., {\bf
165}, 33 (2006).

\bibitem{Pollock} See, e.g., D.D. Pollock, {\it Thermoelectricity: Theory, Thermometry, Tool},
(American Society for Testing and Materials, Philadelphia, PA,
1985).

\bibitem{QPC1}
H. Van-Houten \etal Semicond. Sci. Technol. {\bf 7}, B216 (1992).

\bibitem{Ruitenbeek}
B. Ludoph and J. M. Ruitenbeek, \prb {\bf 59}, 12290 (1999).

\bibitem{QuantumDotsExp}
A. A. M. Staring {\sl et al.},  Europhys. Lett. {\bf 22}, 57 (1993);
L.W. Molenkamp {\sl et al.}, Semicond. Sci. Technol. {\bf 9}, 903
(1994); S. F. Godijn {\sl et al.}, \prl {\bf 82}, 2927 (1999).

\bibitem{Si}
A. I. Hochbaum \etal, Nature {\bf 451}, 163 (2008); A. I. Boukai
\etal Nature {\bf 451}, 168 (2008).


\bibitem{Majumdar}
P. Reddy \etal, Science {\bf 315} 1568 (2007); K. Baheti \etal  Nano
Lett. {\bf 8}, 715 (2008).

\bibitem{Datta}
M. Paulsson and S Datta, \prb {\bf 67}, 241403(R) (2003).


\bibitem{QPCtheory}
A. M. Lunde and K. Flensberg, J. Phys.: Condens. Matter {\bf 17},
3879 (2005); A. M. Lunde {\sl et al.}, \prl {\bf 97}, 256802 (2006).

\bibitem{QDtheory}
C. W. J. Beenakker and A. A. M. Starling, \prb {\bf 46}, 9667
(1992).

\bibitem{MolJuncTHeory}
J. Koch \etal \prb {\bf 70}, 195107 (2004); D. Segal, \prb {\bf 72},
165426 (2005); F. Pauly \etal, cond-mat/0709.3588 (2007).

\bibitem{Freericks}
 J. K. Freericks, V. Zlatic', and A. M. Shvaika, Phys. Rev. B {\bf 75}, 035133
 (2007).

\bibitem{Subroto}
S. Mukerjee, Phys. Rev. B {\bf 72}, 195109 (2005);
 M. R. Peterson \etal, Phys. Rev. B {\bf 76}, 125110 (2007);
 P. Murphy, S. Mukerjee and J. Moore, cond-mat/0805.3374
 (unpublished).

\bibitem{LandauerFormula}
H.- L. Engquist and P. W. Anderson, \prb {\bf 24}, 1151 (1981); U.
Sivan and Y. Imry, \prb {\bf 33}, 552 (1986); P.N. Butcher, J. Phys.
Cond. Matt. {\bf 2}, 4869 (1990).

\bibitem{Ashcroft}
M. Di Ventra, {\it Electrical Transport in Nanoscale Systems},
(Cambridge University Press, 2008).

\bibitem{Cahill}
David G. Cahill {\sl et al.}, App. Phys. Rev. {\bf 93}, 793 (2003).

\bibitem{Fourier}
F. Bonetto \etal  in {\it Mathematical Physics 2000}, A. S. Fokas, A
. Grigoryan, T. Kibble , B. Zegarlinski (Eds.) pp.128 (Imperial
College Press, 2000).

\bibitem{lind} G. Lindblad, Commun. Math. Phys. {\bf 48}, 119 (1976).

\bibitem{vancamp} N. G. Van Kampen, {\it Stochastic Processes in Physics and Chemistry}
(North Holland, Amsterdam, 2001), 2nd ed.

\bibitem{us}
Yu. V. Pershin, Y. Dubi and M. Di Ventra, cond-mat/0803.3216.

\bibitem{spinchain}
A similar form has recently been suggested for spin chains, see C.
Mejia-Monasterio and H. Wichteric, Eur. Phys. J. Special Topics {\bf
151}, 113 (2007).


\bibitem{Max2}
N. Sai \etal \prb {\bf 75}, 115410 (2007).

\bibitem{offdiags}
Note that in a driven system the off-diagonal elements of the DM do
not decay at all, and hence it is crucial to calculate them fully,
as is done in Ref.~\cite{us}. However, in a currentless steady state
they vanish identically. A comparison between the approximate method
and the full many-body calculation for a small system with similar
geometry as discussed here is also presented in Ref.~\cite{us}.

\bibitem{Pothier}
H. Pothier \etal,  Phys. Rev. Lett. {\bf 79}, 3490 (1997).

\bibitem{localT}
A. Nagy \etal, \pra {\bf 53}, 3117 (1996); M. Hartmann \etal \prl
{\bf 93}, 080402 (2004).

\bibitem{Michel}
For a recent review see, e.g., M. Michel \etal,  Int. J. Mod. Phys.
B {\bf 20}, 4855 (2006).

\bibitem{harmonicchain}
C. Gaul and H. B\"{u}ttner, \pre {\bf 76}, 011111 (2007); D. Roy,
\pre {\bf 77}, 062102 (2008).

\bibitem{Chen} Y.-C. Chen \etal, Nano Lett. {\bf 3}, 1691 (2003);
Z. Huang \etal,  Nature Nanotechnology {\bf 2}, 698 (2007).


\bibitem{STDCDTF}
R. D'Agosta and M. Di Ventra, \prl {\bf 98}, 226403 (2007); N.
Bushong and M. Di Ventra, cond-mat/07110762.


\end{thebibliography}
\end{document}